\documentclass[sigconf]{acmart}

\AtBeginDocument{%
  \providecommand\BibTeX{{%
    \normalfont B\kern-0.5em{\scshape i\kern-0.25em b}\kern-0.8em\TeX}}}

\copyrightyear{2020}
\acmYear{2020}
\setcopyright{acmcopyright}\acmConference[CIKM '20]{Proceedings of the 29th ACM International Conference on Information and Knowledge Management}{October 19--23, 2020}{Virtual Event, Ireland}
\acmBooktitle{Proceedings of the 29th ACM International Conference on Information and Knowledge Management (CIKM '20), October 19--23, 2020, Virtual Event, Ireland}
\acmPrice{15.00}
\acmDOI{10.1145/3340531.3411903}
\acmISBN{978-1-4503-6859-9/20/10}

\usepackage{amsmath,amssymb,amsfonts, amsthm}
\usepackage[ruled,vlined,linesnumbered,lined, commentsnumbered]{algorithm2e}
\usepackage{tabularx}
\usepackage{graphicx}
\usepackage{subcaption}
\usepackage[belowskip=2pt,aboveskip=3pt]{caption}
\usepackage{textcomp}
\usepackage{url}
\usepackage{booktabs}
\usepackage{epstopdf}
\usepackage{multirow}
\usepackage{array}
\usepackage{enumitem}
\usepackage{wrapfig}
\usepackage{tcolorbox}
\usepackage{bbm}
\usepackage{bm}
\usepackage{hyperref}
\usepackage{lipsum}
\usepackage[noend]{algpseudocode}

\theoremstyle{definition}
\newtheorem{definition}{Definition}[section]

\begin{document}

% \title{CARE-GNN: Camouflage Resistant Graph Neural Network for Fraud Detection}
\title{Enhancing Graph Neural Network-based Fraud Detectors against Camouflaged Fraudsters}

% \author{Anonymous Author(s)}

\author{Yingtong Dou$^{1}$, Zhiwei Liu$^{1}$, Li Sun$^{2}$, Yutong Deng$^{2}$, Hao Peng$^{3}$, Philip S. Yu$^{1}$}
\affiliation{
\institution{
$^1$Department of Computer Science, University of Illinois at Chicago\\
$^2$School of Computer Science, Beijing University of Posts and Telecommunications\\
$^3$Beijing Advanced Innovation Center for Big Data and Brain Computing, Beihang University\\
}
}
\email{{ydou5, zliu213, psyu}@uic.edu, {l.sun, buptdyt}@bupt.edu.cn, penghao@act.buaa.edu.cn}

% \affiliation{%

\renewcommand{\shortauthors}{Dou and Liu, et al.}

\begin{abstract}
% Recently, Graph Neural Networks (GNNs) have been widely applied to fraud detection problems where fraudsters usually reveal suspicious signals after aggregating their neighborhood information from different relations.
Graph Neural Networks (GNNs) have been widely applied to fraud detection problems in recent years, revealing the suspiciousness of nodes by aggregating their neighborhood information via different relations.
However, few prior works have noticed the camouflage behavior of fraudsters, which could hamper the performance of GNN-based fraud detectors during the aggregation process.  
In this paper, we introduce two types of camouflages based on recent empirical studies, i.e., the feature camouflage and the relation camouflage.
Existing GNNs have not addressed these two camouflages, which results in their poor performance in fraud detection problems.
Alternatively, we propose a new model named CAmouflage-REsistant GNN (CARE-GNN), to enhance the GNN aggregation process with three unique modules against camouflages.
Concretely, we first devise a label-aware similarity measure to find informative neighboring nodes.
Then, we leverage reinforcement learning (RL) to find the optimal amounts of neighbors to be selected.
Finally, the selected neighbors across different relations are aggregated together.
Comprehensive experiments on two real-world fraud datasets demonstrate the effectiveness of the RL algorithm.
The proposed CARE-GNN also outperforms state-of-the-art GNNs and GNN-based fraud detectors. We integrate all GNN-based fraud detectors as an opensource toolbox\footnote{\url{https://github.com/safe-graph/DGFraud}}. The CARE-GNN code and datasets are available at \url{https://github.com/YingtongDou/CARE-GNN}. 
% CARE-GNN exhibits stable performance under various experimental settings.
% Meanwhile, the $\mathsf{DGF}$ has the best detection performance among a wide spectrum of baselines including shallow/deep node embedding algorithms and the state-of-the-art GNN-based fraud detection methods. 
% It also exhibits a good performance with only small amount of training data (1\%) which alleviates the costly labelling problem. 
% Comparing to previous fraud detection methods, $\mathsf{DGF}$ retains the end-to-end fashion of GNN and makes it more flexible to other fraud detection scenarios. It also sheds light on improving GNNs via neighborhood routing.
\end{abstract}

\begin{CCSXML}
<ccs2012>
   <concept>
       <concept_id>10002978.10003022.10003026</concept_id>
       <concept_desc>Security and privacy~Web application security</concept_desc>
       <concept_significance>500</concept_significance>
       </concept>
   <concept>
       <concept_id>10010147.10010257.10010293.10010294</concept_id>
       <concept_desc>Computing methodologies~Neural networks</concept_desc>
       <concept_significance>500</concept_significance>
       </concept>
 </ccs2012>
\end{CCSXML}

\ccsdesc[500]{Security and privacy~Web application security}
\ccsdesc[500]{Computing methodologies~Neural networks}

\keywords{Graph Neural Networks, Fraud Detection, Reinforcement Learning}

\maketitle

\section{Introduction}
\label{sec01:intro}

% Global background and meaning
% Summarize the weaknesses of existing methods
% our work and motivation

As Internet services thrive, they also incubate various kinds of fraudulent activities~\cite{jiang2016suspicious}.
% In order to disseminate disinformation or steal money from online business, fraudsters hack systems directly or disguise as regular users to bypass the anti-fraud system.
Fraudsters disguise as regular users to bypass the anti-fraud system and disperse disinformation~\cite{yang2020rumor} or reap end-users' privacy~\cite{sun2020kollector}.
To detect those fraudulent activities, graph-based methods have become an effective approach in both academic~\cite{li2020flowscope, wang2018deep, dhawan2019spotting} and industrial communities~\cite{breuer2020friend, nilforoshan2019silcendice, zhong2020financial}.
% Graph-based methods connect entities with relations and reveal the suspiciousness of these entities at the graph level, since fraudsters with the same goal usually have explicit or latent connections~\cite{Koutra2011,akoglu2015graph}.
Graph-based methods connect entities with different relations and reveal the suspiciousness of these entities at the graph level, since fraudsters with the same goal tend to connect with each other~\cite{akoglu2015graph}.

% \textcolor{red}{(Question: Do we need this paragraph?)} To spot the fraudsters via graph information, traditional approaches detect suspicious dense blocks in the graph~\cite{Shin2017, dhawan2019spotting, li2020flowscope, wang2018deep} or pass messages between connected nodes to infer the node suspiciousness~\cite{Rayana2015, Wang2017, kaghazgaran2018combating, breuer2020friend}. 
% Some works~\cite{kumar2018rev2, nilforoshan2019silcendice} devise novel metrics to measure node suspiciousness and other works~\cite{hu2019cash, zhong2020financial} adopt heterogeneous information networks~\cite{wang2019heterogeneous} to model the entities.

Recently, as the development of Graph Neural Networks (GNNs) (e.g., GCN~\cite{kipf2016semi}, GAT~\cite{velivckovic2017graph}, and GraphSAGE~\cite{hamilton2017inductive}), 
many GNN-based fraud detectors have been proposed to detect opinion fraud~\cite{wang2019fdgars, li2019spam, liu2020alleviating}, financial fraud~\cite{liu2019geniepath, liu2018heterogeneous, wang2019semi}, mobile fraud~\cite{wen2020adversary}, and cyber criminal~\cite{zhang2019key}. In contrast to traditional graph-based approaches, GNN-based methods aggregate neighborhood information to learn the representation of a center node with neural modules.
They can be trained in an \textit{end-to-end} and \textit{semi-supervised} fashion, which saves much feature engineering and data annotation cost.
% 1) GNN-based methods enjoy the \textit{end-to-end} learning fashion which means they only take graph information as input and output the suspicious estimation of nodes.
% It saves much feature engineering cost in practice.
% 2) Many GNNs are capable of \textit{semi-supervised learning} (SSL) where unlabeled nodes also attend the training process and provide additional information.
% It fits the fraud detection scenario very well since the data annotation is costly.
% 3) GNN could jointly learn various types of edges and features and encode them into a unified representation space. 

% However, existing GNN-based fraud detection works only apply GNNs to specific applications but ignore the camouflage behaviors of fraudsters,
However, existing GNN-based fraud detection works only apply GNNs in a narrow scope while ignoring the camouflage behaviors of fraudsters,
which have been drawing great attention from both researchers~\cite{zheng2017smoke, kaghazgaran2018combating, kaghazgaran2019wide, dou2020robust} and practitioners~\cite{li2019spam, wen2020adversary, breuer2020friend}.
Meanwhile, theoretical studies prove the limitations and vulnerabilities of GNNs when graphs have noisy nodes and edges~\cite{chen2020smooth, hou2020measure, chen2019label, sun2018adversarial}.
Therefore, failing to tackle the camouflaged fraudsters would sabotage the performance of GNN-based fraud detectors.
Though some recent works~\cite{chen2019label, liu2020alleviating, wen2020adversary, hou2020measure, franceschi2019learning} have noticed similar challenges,
their solutions either fail to fit the fraud detection problems or break the end-to-end learning fashion of GNNs.

\begin{figure}
    \centering
    \includegraphics[width=0.8\linewidth]{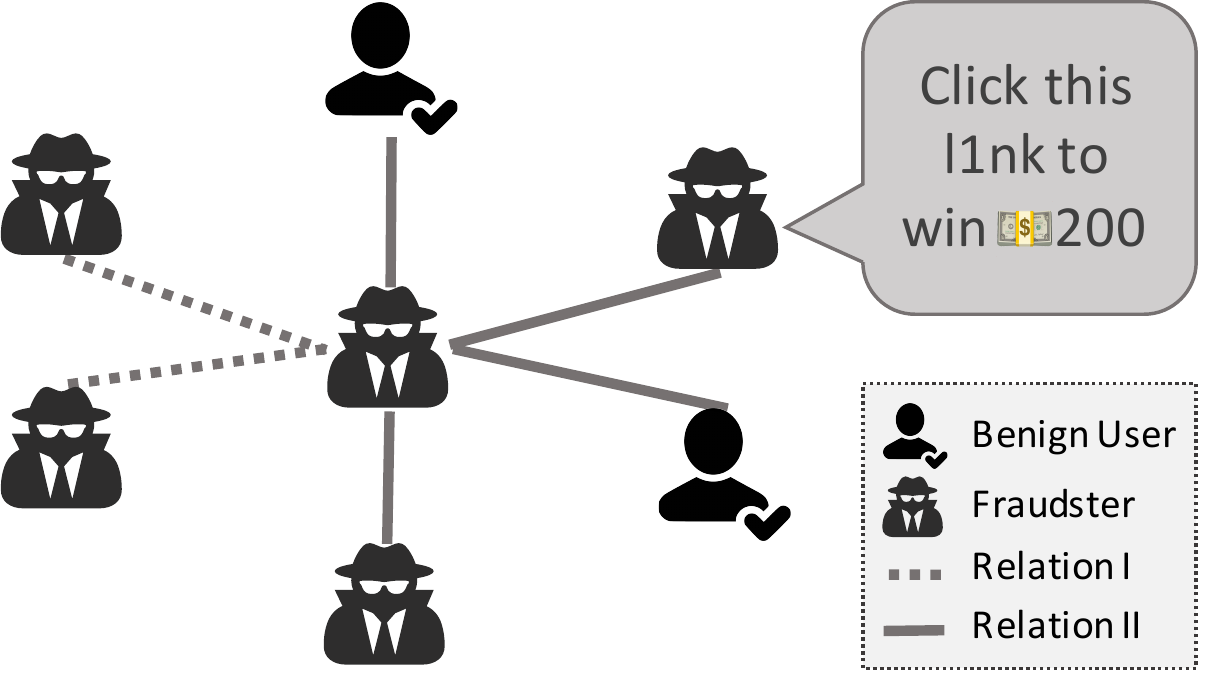}
    \caption{Two types of fraudster camouflage. (1) Feature camouflage: fraudsters add special characters to the text and make it delusive for feature-based spam detectors. 
    (2) Relation camouflage: center fraudster connects to many benign entities under Relation II to attenuate its suspiciousness.}
    \label{fig:camo}
\end{figure}

To demonstrate the challenges induced by camouflaged fraudsters during the neighbor aggregation of GNNs, as shown in Figure~\ref{fig:camo}, we construct a graph with two relations and two types of entities. The relation can be any common attributes supposing to be shared by similar entities (e.g., the \textit{User-IP-User} relation connects entities with the same IP address). 
% We employ the GNN to predict the suspiciousness of the center entity.
% From the perspective of fraudsters, we introduce two types of \textit{camouflages} as follows:
There are two types of \textit{camouflages} as follows:
\textbf{1) Feature camouflage:}
% recent studies show that smart fraudsters could adjust their behaviors~\cite{ge2018securing}, add special characters in reviews~\cite{li2019spam, wen2020adversary},
smart fraudsters may adjust their behaviors~\cite{ge2018securing, dou2020robust}, add special characters in reviews~\cite{li2019spam, wen2020adversary} (so-called spamouflage),
or employ deep language generation models~\cite{kaghazgaran2019wide} to gloss over explicit suspicious outcomes.
% or employ advanced deep language generation models~\cite{kaghazgaran2019wide} to evade the detectors using behavior/language features.
% Like Figure~\ref{fig:camo} shows, a fraudster can add some special characters to a fake review which could bypass semantic-based features ~\cite{wen2020adversary}.
Like Figure~\ref{fig:camo} shows, a fraudster can add some special characters to a fake review, which helps to bypass feature-based detectors~\cite{wen2020adversary}.
\textbf{2) Relation camouflage:}
previous works~\cite{zheng2017smoke, kaghazgaran2018combating} show that crowd workers are actively committing opinion fraud on online social networks.
They can probe the graphs used by defenders~\cite{yang2020secure} and adjust their behavior to alleviate the suspiciousness~\cite{yang2020rumor}.
Specifically, these crafty fraudsters camouflage themselves via connecting to many benign entities (i.e., posting regular reviews or connecting to reputable users).
As Figure~\ref{fig:camo} shows, under Relation II, there are more benign entities than fraudsters.

% \noindent \textbf{(3) Inter-relation camouflage.}
% Professional fraudsters can infiltrate the detection system~\cite{ge2018securing, sun2018adversarial} or leverage open-source resources~\cite{liu2020alleviating} to infer the detector configurations.
% As Figure~\ref{fig:camo} shows, Relation I would get higher aggregation weight than Relation II during GNN training since there are more fraudsters under Relation II.
% If the center fraudster figures it out and switch its connection to Relation II, since Relation II has a smaller weight, the center fraudster could lower its suspiciousness after aggregating all neighbors.  

Directly applying GNNs to graphs with camouflaged fraudsters will hamper the neighbor aggregation process of GNNs.
As Figure~\ref{fig:camo} shows,
if we aggregate neighbors with the intriguing reviews as node features, it will probably smooth out the suspiciousness of the center fraudster~\cite{liu2020alleviating, hou2020measure}.
Similarly, if we aggregate all neighbors under Relation II, where there are more dissimilar neighbors, it will eliminate the suspiciousness of the center fraudster.

Considering the agility of real-world fraudsters~\cite{ge2018securing, dou2020robust}, designing GNN-based detectors that exactly capture these camouflaged fraudsters is impractical.
Therefore, based on the outcomes of two camouflages and the aggregation process of GNNs, we propose \textit{three} neural modules to enhance the GNNs against the camouflages.
\textbf{1)} For the feature camouflage, we propose a
\textbf{label-aware similarity measure} to find the most similar neighbors based on node features.
Specifically, we design a neural classifier as a similarity measure,
which is directly optimized according to experts with domain knowledge (i.e., annotated data).
% For the relation camouflage, we devise a \textbf{similarity-aware neighbor selector} to select the similar neighbors as many as possible under a relation.
\textbf{2)} For the relation camouflage, we devise a \textbf{similarity-aware neighbor selector} to select the similar neighbors of a center node within a relation.
Furthermore, we leverage reinforcement learning (RL) to adaptively find the optimal neighbor selection threshold along with the GNN training process.
\textbf{3)} We utilize the neighbor filtering thresholds learned by RL to formulate a \textbf{relation-aware neighbor aggregator} which combines neighborhood information from different relations and obtains the final center node representation.     

We integrate above three modules together with general GNN frameworks and name our model as \underline{CA}mouflage \underline{RE}sistant \underline{G}raph \underline{N}eural \underline{N}etwork (CARE-GNN).
Experimental results on two real-world fraud datasets demonstrate that our model boosts the GNN performance on graphs with camouflaged fraudsters.
The proposed neighbor selector can find optimal neighbors and CARE-GNN outperforms state-of-the-art baselines under various settings.

We highlight the advantages of CARE-GNN as follows:
\begin{itemize}
    \item \textbf{Adaptability.} CARE-GNN adaptively selects best neighbors for aggregation given arbitrary multi-relation graph.
   
    \item \textbf{High-efficiency.} CARE-GNN has a high computational efficiency without attention and deep reinforcement learning. 
   
    \item \textbf{Flexibility.} Many other neural modules and external knowledge can be plugged into the CARE-GNN.

\end{itemize}
\begin{table}
\caption{Glossary of Notations.}  
\resizebox{\linewidth}{!}{%
\begin{tabular}{r|l}  
\hline\hline
\textbf{Symbol} & \textbf{Definition} \\
\hline
$\mathcal{G}; \mathcal{V}; \mathcal{E}; \mathcal{X}$ & Graph; Node set; Edge set; Node feature set \\
\hline
$y_{v}; Y$ & Label for node $v$; Node label set \\
\hline
$r; R$ & Relation; Total number of relations\\
\hline
$l; L$ & GNN layer number; Total number of layers \\
\hline
$b; B$ & Training batch number; Total number of batches \\
\hline
$e; E$ & Training epoch number; Total number of epochs \\
\hline
$\mathcal{V}_{train}; \mathcal{V}_{b}$ & Nodes in the training set; Node set at batch $b$ \\
\hline
$\mathcal{E}^{(l)}_{r}$ & Edge set under relation $r$ at the $l$-th layer \\
\hline
$\mathbf{h}_{v}^{(l)}$ & The embedding of node $v$ at the $l$-th layer  \\
\hline
$\mathbf{h}_{v, r}^{(l)}$ &  The embedding of node $v$ under relation $r$ at the $l$-th layer     \\
\hline
$\mathcal{D}^{(l)}(v, v^{\prime})$ & The distance between node $v$ and $v^{\prime}$ at the $l$-th layer \\
\hline
$S^{(l)}(v, v^{\prime})$ &  The similarity between node $v$ and $v^{\prime}$ at the $l$-th layer\\ 
\hline
$p_{r}^{(l)}\in P$   &  The filtering threshold for relation $r$ at the $l$-th layer  \\
\hline
$a_{r}^{(l)} \in A; \tau $ & RL action space; Action step size  \\
\hline
$G(\mathcal{D}_{r}^{(l)})$ & Average neighbor distances for relation $r$ at the $l$-th layer \\
\hline
$f(p_{r}^{(l)}, a_{r}^{(l)})$ &  RL reward function\\
\hline
$\textnormal{AGG}_{r}^{(l)}$ &  Intra-relation aggregator for relation $r$ at the $l$-th layer           \\
\hline
$\textnormal{AGG}_{all}^{(l)}$ & Inter-relation aggregator at the $l$-th layer \\
\hline
$\mathbf{z}_{v}$ & Final embedding for node $v$         \\
% \hline
% $\lambda$ & Loss weight for the similarity measure        \\
\hline
\hline
\end{tabular}}
\label{tab:notation}
\end{table}

\section{Problem Definition}
\label{sec03:prelim}

In this section, we first define the multi-relation graph and the graph-based fraud detection problem.
Then, we introduce how to apply GNN to fraud detection problems. All important notations in this paper are summarized in Table~\ref{tab:notation}.

\definition 
\textbf{Multi-relation Graph.}
We define a multi-relation graph as $\mathcal{G}=\left\{\mathcal{V}, \mathcal{X}, \{\mathcal{E}_{r}\}|_{r=1}^{R}, Y\right\}$, where $\mathcal{V}$ is the set of nodes $\{v_{1}, \dots, v_{n}\}$.
Each node $v_{i}$ has a $d$-dimensional feature vector $\mathbf{x}_{i}\in\mathbb{R}^{d}$ and $\mathcal{X} = \{\mathbf{x}_{1}, \dots, \mathbf{x}_{n}\}$ represents a set of all node features.
$e_{i, j}^{r} = (v_{i}, v_{j})\in \mathcal{E}_{r}$ is an edge between $v_{i}$ and $v_{j}$ with a relation $r \in \{1, \cdots, R\} $.
Note that an edge can be associated with multiple relations and there are $R$ different types of relations.
$Y$ is the a set of labels for each node in $\mathcal{V}$.
% $\mathbf{A} \in \mathbb{R}^{n\times n}$ is the adjacency matrix of $\mathcal{G}$, where $\mathbf{A}_{i, j} = 1$ if $(v_{i}, v_{j}) \in \mathcal{E}$. 

\begin{figure*}
    \centering
   \includegraphics[width=0.95\textwidth]{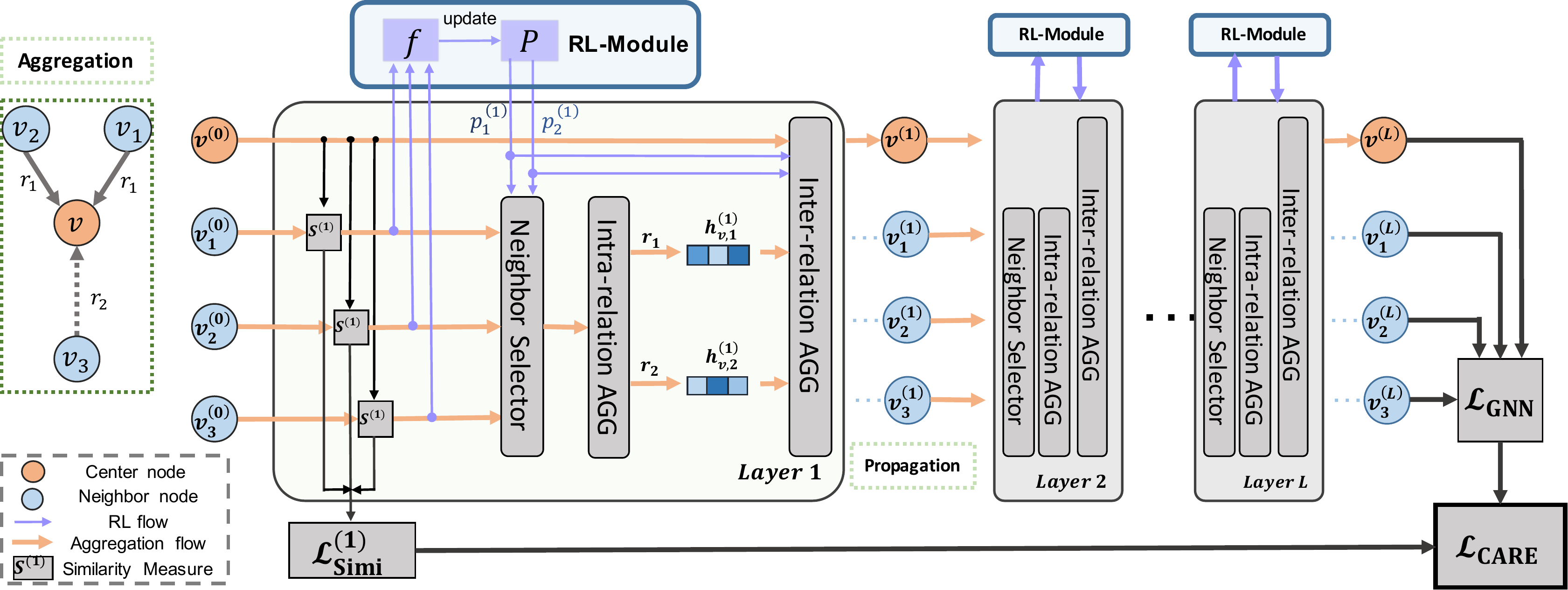}
    \caption{The aggregation process of proposed CARE-GNN at the training phase.}
    \label{fig:framework}
\end{figure*}

\definition 
\label{def:multi-relation}
\textbf{Fraud Detection on Graph.}
For the fraud detection problem, the node $v$ represents the target entity whose suspiciousness needs to be justified.
For example, it can be a review on the review website~\cite{li2019spam, Rayana2015} or a transaction in the trading system~\cite{liu2019geniepath, wang2019semi}.
The node has a label $y_{v}\in\{0, 1\}\in Y$ where 0 represents \textit{benign} and 1 represents \textit{suspicious}.
The relations $R$ are rules, interactions, or shared attributes between nodes, e.g., two reviews from the same user~\cite{liu2020alleviating} or transactions from the same devices~\cite{liu2018heterogeneous}.
The graph-based fraud detection problem is a semi-supervised binary node classification problem on the graph.
Graph-based fraud detectors are trained based on the labeled node information along with the graph composed of multi-relations.
The trained models are then used to predict the suspiciousness of unlabeled nodes.
% It can be an transductive or inductive learning problem dependent on whether the testing nodes attend the training process.

\definition 
\textbf{GNN-based Fraud Detection.}
% given the notations to outline the aggregator
A Graph Neural Network (GNN) is a deep learning framework to embed graph-structured data via aggregating the information from its neighboring nodes~\cite{kipf2016semi,velivckovic2017graph,hamilton2017inductive}. Based on the defined multi-relation graph in Definition~\ref{def:multi-relation}, we unify the formulation of GNNs from the perspective of neighbor aggregation (as shown in the left side of Figure~\ref{fig:framework}):
\begin{equation}\label{eq:gnn_framework}
    \mathbf{h}_{v}^{(l)}= \sigma\left(\mathbf{h}_{v}^{(l-1)} \oplus \operatorname{AGG}^{(l)}\left(\{\mathbf{h}_{v^{\prime}, r}^{(l-1)}: (v,v^{\prime}) \in \mathcal{E}^{(l)}_{r}\}|_{r=1}^{R}\right)\right).
\end{equation}
For a center node $v$, $\mathbf{h}_{v}^{(l)}$ is the hidden embedding at $l$-th layer and $\mathbf{h}_{v}^{(0)} = \mathbf{x}_{i}$ is the input feature.
$\mathcal{E}^{(l)}_{r}$ denotes edges under relation $r$ at the $l$-th layer. $\mathbf{h}_{v^{\prime}, r}^{(l-1)}$ is the embedding of neighboring node $v^{\prime}$ under relation $r$.
$\operatorname{AGG}$ represents the aggregation function that mapping the neighborhood information from different relations into a vector, e.g., mean aggregation~\cite{hamilton2017inductive} and attention aggregation~\cite{velivckovic2017graph}.
$\oplus$ is the operator that combines the information of $v$ and its neighboring information, e.g., concatenation or summation~\cite{hamilton2017inductive}.

For fraud detection problems, we first construct a multi-relation graph based on domain knowledge.
Then, the GNN is trained with partially labeled nodes supervised by binary classification loss functions.
Instead of directly aggregating the neighbors for all relations, we separate the aggregation part as \textit{intra-relation} aggregation and \textit{inter-relation} aggregation process.
During the intra-relation aggregation process, the embedding of neighbors under each relation is aggregated simultaneously.
Then, the embeddings for each relation are combined during the inter-relation aggregation process.
Finally, the node embeddings at the last layer are used for prediction. 
% \textcolor{red}{add intra-inter introduction} 

% To avoid the confusion of terms, we use \textit{graph} to represent the nodes, edges and their attributes and use \textit{network} to represent neural networks. We interchangeably use \textit{neighboring nodes}, \textit{neighbors}, and \textit{neighborhood} to represent the nodes 1-hop or 2-hop away from the center node. 

\section{Proposed Model}
\label{sec03:method}

\subsection{Model Overview}
\label{sec04:overview}

The proposed CARE-GNN has three neural modules and its pipeline is shown in Figure~\ref{fig:framework}.
For a center node $v$, we first compute its neighbor similarities based with proposed label-aware similarity measure (Section~\ref{sec04:similarity}).
Then we filter the dissimilar neighbors under each relation with the proposed neighbor selector (Section~\ref{sec04:threshold}). 
The neighbor selector is optimized using reinforcement learning during training the GNN (purple module in Figure~\ref{fig:framework}).
At the aggregation step, we first use the intra-relation aggregator to aggregate neighbor embeddings under each relation.
Then, we combine embeddings across different relations with the inter-relation aggregator (Section~\ref{sec04:aggregation}). The optimization steps and the algorithm procedure are presented in Section~\ref{sec04:optimization} and Algorithm~\ref{alg:care-gnn}, respectively.
% \textcolor{red}{add highlight.}
% users to filter the irrelevant (i.e., dissimilar) neighbors using the proposed similarity measure in Eq.~(\ref{eq:similarity}) before applying GNNs.
% Then the GNN is directly applied to an optimal graph with informative neighbors. 
% Different from previous works which insert new layers (e.g., similarity, pooling) between regular GNN layers and train GNNs with a full batch, we adopt the mini-batch training fashion of GraphSAGE~\cite{hamilton2017inductive}. 

\subsection{Label-aware Similarity Measure}
\label{sec04:similarity}

Previous studies have introduced various fraudster camouflage types from behavior~\cite{ge2018securing, dou2020robust} and semantic~\cite{kaghazgaran2019wide, wen2020adversary} perspectives.
Those camouflages could make the features of fraudsters and benign entities similar to each other, and further mislead GNNs to generate uninformative node embeddings.
To tackle those node feature camouflages, we deem that an effective similarity measure is needed to filter the camouflaged neighbors before applying GNNs.
Previous works have proposed unsupervised similarity metrics like \textit{Cosine Similarity}~\cite{liu2020alleviating} or Neural Networks~\cite{yilmaz2020unsupervised}.
However, many fraud problems like financial fraud and opinion fraud require extra domain knowledge to identify fraud instances.
For example, in opinion fraud, unsupervised similarity measures could not identify the camouflaged fake reviews, which are even indistinguishable by humans ~\cite{kaghazgaran2019wide}.
Therefore, we need a parameterized similarity measure to compute node similarity with supervised signals from domain experts (e.g., high-fidelity data annotations).
% This similarity measure should be part of the GNN pipeline which is jointly trained with GNNs.
% So it could adaptively learn the similarities among raw features or embeddings at each GNN layer.
% In this way, we could retain the end-to-end learning fashion of GNNs and save much costs in devising and updating handcrafted similarity metrics. 

For the parameterized similarity measure, AGCN~\cite{li2018adaptive} employs a \textit{Mahalanobis distance} plus a \textit{Gaussian kernel}, and DIAL-GNN~\cite{chen2019deep} uses the parameterized \textit{cosine similarity}.
However, those two types of measures suffer from high time complexity $O(|\mathcal{V}|\Bar{D}d)$, where $\Bar{D}$ is the average degree of nodes which is extremely high in real-world graphs (see Table~\ref{tab:stat}) and $d$ is the feature dimension.

\vspace{1mm}
\noindent \textbf{Label-aware Similarity Measure.} 
Inspired by LAGCN~\cite{chen2019label} which uses a Multi-layer Perceptron (MLP) as the edge label predictor, we employ a one-layer MLP as the node label predictor at each layer and use the $l_{1}$-distance between the prediction results of two nodes as their similarity measure.
For a center node $v$ under relation $r$ at the $l$-th layer and edge $(v, v^{\prime})\in \mathcal{E}_{r}^{(l-1)}$, the distance between $v$ and $v^{\prime}$ is the $l_1$-distance of two embeddings:
\begin{equation}
    \mathcal{D}^{(l)}(v, v^{\prime}) = \left\|\sigma\left(MLP^{(l)}(\mathbf{h}^{(l-1)}_{v})\right) - \sigma\left(MLP^{(l)}(\mathbf{h}^{(l-1)}_{v^{\prime}})\right)\right\|_{1},
    \label{eq:distance}
\end{equation}
and we can define the similarity measure as:
\begin{equation}
    S^{(l)}(v, v^{\prime}) = 1 - \mathcal{D}^{(l)}(v, v^{\prime}), \label{eq:similarity}
\end{equation}
where each layer has its own similarity measure.
The input of MLP at the $l$-th layer is the node embedding at the previous layer, and the output of MLP is a scalar which is then fed into a non-linear activation function $\sigma$ (we use \texttt{tanh} in our work).
To save the computational cost, we only take the embedding of the node itself as the input instead of using combined embeddings like the LAGCN~\cite{chen2019label}.
Therefore, taking the $S_{r}^{(1)}(v, v^{\prime})$ as an example where the input is the raw feature, the time complexity of the proposed similarity measure reduces significantly from $O(|\mathcal{V}|\Bar{D}d)$ to $O(|\mathcal{V}|d)$ since it predicts the node label solely based on its feature. 

\vspace{1mm}
\noindent \textbf{Optimization.} To train the similarity measure together with GNNs, a heuristic approach is to append it as a new layer before the aggregation layer of GCN~\cite{li2018adaptive}.
However, if the similarity measure could not effectively filter the camouflaged neighbors at the first layer, it will hamper the performance of following GNN layers.
Consequently, the MLP parameters cannot be well-updated through the back-propagation process.
To train the similar measure with a direct supervised signal from labels, like~\cite{verma2019graphmix}, we define the cross-entropy loss of the MLP at $l$-layer as:
\begin{equation}\label{eq:simi_loss}
    \mathcal{L}^{(l)}_{\textnormal{Simi}} = \sum_{v\in \mathcal{V}}-\log\left(y_{v}\cdot\sigma\left(MLP^{(l)}(\mathbf{h}^{(l)}_{v})\right)\right).
\end{equation}
During the training process, the similarity measure parameters are directly updated through the above loss function.
It guarantees similar neighbors can be quickly selected within the first few batches and help regularize the GNN training process.

\subsection{Similarity-aware Neighbor Selector}
\label{sec04:threshold}

Given the similarity scores between the center node and its neighbors with Eq.~(\ref{eq:similarity}), we should select similar neighbors (i.e., filter camouflaged ones) to improve the capability of GNNs.
%Some previous methods~\cite{chen2019deep, liu2020alleviating} rank neighbors based on their importance scores and empirically find a fixed threshold to filter less important ones. Other works employ importance sampling techniques to sample the important neighbors~\cite{chen2018fastgcn, chen2019label, zou2019layer}.
According to the relation camouflage, fraudsters may connect to different amounts of benign entities under different relations~\cite{yang2020rumor}.
However, since data annotation is costly for real-world fraud detection problems, computing the number of similar neighbors under each relation through data labeling is impossible.    
We should devise an adaptive filtering/sampling criteria to select an optimal amount of similar neighbors automatically.
% Under the same relation, for two center nodes with opposite classes, we suppose that each node class has its own percentage of similar neighbors.
Thus, we design a similarity-aware neighbor selector.
It selects similar neighbors under each relation using \textit{top-p} sampling with an adaptive filtering threshold.
We also devise a reinforcement learning (RL) algorithm to find optimal thresholds during the GNN training process.

\subsubsection{Top-p Sampling}
\label{sec04:sampling}
We employ \textit{top-p} sampling to filter camouflaged neighbors under each relation.
The filtering threshold for relation $r$ at the $l$-th layer is $p_{r}^{(l)}\in[0,1]$.
The closed interval means we could discard or keep all neighbors of a node under a relation.
Specifically, during the training phase, for a node $v$ in current batch under relation $r$, we first compute a set of similarity scores $\{S^{(l)}(v, v^{\prime})\}$ using Eq.~(\ref{eq:similarity}) at the $l$-th layer where $(v,v^{\prime})\in \mathcal{E}_{r}^{(l)}$.
$\mathcal{E}_{r}^{(l)}$ is a set of edges under relation $r$ at the $l$-th layer.
Then we rank its neighbors based on $\{S^{(l)}(v, v^{\prime})\}$ in descending order and take the first $p_{r}^{(l)}\cdot|\{S^{(l)}(v, v^{\prime})\}|$ neighbors as the selected neighbors at the $l$-th layer. 
All other nodes are discarded at the current batch and will not attend the aggregation process.
The \textit{top-p} sampling process is applied to the center node at every layer for each relation. 
% $P$ represents a set of filtering threshold used in our model.

\subsubsection{Finding the Optimal Thresholds with RL}\label{sec04:RL}
Previous works~\cite{chen2019deep, liu2020alleviating} set the filtering threshold as a hyperparameter and tune it with validation to find the optimal value.
However, their models are built upon homogeneous benchmark graphs, and without noise induced by camouflaged fraudsters.
However, owing to the multi-relation graph of fraud problems as well as the relation camouflage problem, we need an automatic approach to find the optimal threshold $p_{r}^{(l)}$ for each relation.
Since $p_{r}^{(l)}$ is a probability and has no gradient, we cannot use back-propagation from the classification loss to update it.
Meanwhile, given a $p_{r}^{(l)}$, it is infeasible to estimate the quality of selected neighbors solely based on the similarity scores under the current batch/epoch.
To overcome the above challenges, we propose to employ a reinforcement learning (RL) framework to find optimal thresholds.

Concretely, we formulate the RL process as a Bernoulli Multi-armed Bandit (BMAB) $\mathcal{B}(A, f, T)$ between the neighbor selector and the GNN with the similarity measure. $A$ is the action space, $f$ is the reward function, and $T$ is the terminal condition~\cite{vermorel2005multi}. Given an initial $p_{r}^{(l)}$, the neighbor selector choose to increase or decrease $p_{r}^{(l)}$ as actions and the reward is dependent on the average distance differences between two consecutive epochs.
Next, we introduce the details of each BMAB component:

\begin{itemize}[leftmargin=*]
    % \item \textbf{State.} The state $s_{b}^{(l)}$ at batch $b$ and the $l$-th layer is represented by a set of similarity scores, where $v\in \mathcal{V}_{b}$:
    % \begin{equation}\label{eq:state}
    %     s_{b}^{(l)} = \{S^{(l)}(v, v^{\prime}): (v,v^{\prime}) \in \mathcal{E}^{(l)}_{r}\}|_{r=1}^{R}.
    % \end{equation}
    
    \item \textbf{Action.} The action represents how RL updates the $p_{r}^{(l)}$ based on the reward. Since $p_{r}^{(l)}\in[0,1]$, we define the action $a_{r}^{(l)}$ as plus or minus a fixed small value $\tau \in [0,1]$ from $p_{r}^{(l)}$.
    
    \item \textbf{Reward.} The optimal $p_{r}^{(l)}$ is expected to find the most similar (i.e., minimum distances in Eq. (\ref{eq:distance})) neighbors of a center node under relation $r$ at the $l$-th layer.
    We cannot sense the state of GNN due to its black-box nature; thus, we design a binary stochastic reward solely based on the average distance differences between two consecutive epochs.
    The average neighbor distances for relation $r$ at the $l$-th layer for epoch $e$ is:
    \begin{equation}
    G(\mathcal{D}_{r}^{(l)})^{(e)} = \frac{\sum_{v\in\mathcal{V}_{train}} \mathcal{D}_{r}^{(l)}(v, v^{\prime})^{(e)}}{|\mathcal{V}_{train}|}.
    \label{eq:avg_distance}
    \end{equation}
    Then, we can define the reward for epoch $e$ as:
    \begin{equation}
    f(p_{r}^{(l)}, a_{r}^{(l)})^{(e)} = \left\{\begin{array}{l}+1, G(\mathcal{D}_{r}^{(l)})^{(e-1)} - G(\mathcal{D}_{r}^{(l)})^{(e)} \geq 0, \\ -1, G(\mathcal{D}_{r}^{(l)})^{(e-1)} - G(\mathcal{D}_{r}^{(l)})^{(e)} < 0. \end{array}\right.
    \label{eq:threshold_rewards}
    \end{equation}
    The reward is positive when the average distance of newly selected neighbors at epoch $e$ is less than that of the previous epoch, and vice versa.
    It is not easy to estimate the cumulative reward. Thus, we use the immediate reward to update the action greedily without exploration.
    Concretely, we increase $p_{r}^{(l)}$ with a positive reward and decrease it vice versa.
    
    \item \textbf{Terminal.} We define the terminal condition for RL as:
    \begin{equation}\label{eq:terminal}
      \big|\sum_{e-10}^{e} f(p_{r}^{(l)}, a_{r}^{(l)})^{(e)}\big| \leq 2, \;where \; e \geq 10.
    \end{equation}
    It means that the RL converges in the recent ten epochs and indicates an optimal threshold $p_{r}^{(l)}$ is discovered. 
    After the RL module terminates, the filtering thresholds are fixed as the optimal one until the convergence of GNN.
\end{itemize}

\vspace{1mm}
\noindent \textbf{Discussion.} Different node classes may have different amounts of similar neighbors under the same relation.
For instance, as Table~\ref{tab:stat} shows, under the \textit{R-S-R} relation of the Yelp dataset, for positive nodes, only 5\% of their neighbors have the same label.
This is due to the class-imbalance nature of fraud problems and the relation camouflage of fraudsters.
According to the cost-sensitive learning research~\cite{sahin2013cost}, misclassifying a fraudster has a much higher cost to defenders than misclassifying a benign entity.
Meanwhile, a large number of benign entities already fuel sufficient information for the classifier.
Therefore, to accelerate the training process, we compute the filtering thresholds by only considering positive center nodes (i.e., fraudsters) and apply them for all node classes. 
The complete RL process is shown in Lines 15-19 of Algorithm~\ref{alg:care-gnn}.  
The experiment results in Section~\ref{sec05:training_analysis} verify the RL effectiveness.

\subsection{Relation-aware Neighbor Aggregator}
\label{sec04:aggregation}

After filtering neighbors under each relation, the next step is to aggregate the neighbor information from different relations.
Previous methods adopt attention mechanism~\cite{liu2019geniepath, zhang2019key, wang2019semi} or devise weighting parameters~\cite{liu2018heterogeneous} to learn the relation weights during aggregating information from different relations.
However, supposing we have selected the most similar neighbors under each relation, the attention coefficients or weighting parameters should be similar among different relations.
Thus, to save the computational cost while retaining the relation importance information, we directly apply the optimal filtering threshold $p_{r}^{(l)}$ learned by the RL process as the inter-relation aggregation weights.
Formally, under relation $r$ at the $l$-th layer, after applying the \textit{top-p} sampling, for node $v$, we define the \textbf{intra-relation} neighbor aggregation as follows: 
\begin{equation}\label{eq:aggregator}
    \mathbf{h}_{v, r}^{(l)}= \textnormal{ReLU}\left(\textnormal{AGG}_{r}^{(l)}\left(\left\{ \mathbf{h}_{v^{\prime}}^{(l-1)}: (v,v^{\prime}) \in \mathcal{E}_{r}^{(l)}\right\}\right)\right),
\end{equation}
where a mean aggregator is used for all $\textnormal{AGG}_{r}^{(l)}$.
Then, we define the \textbf{inter-relation} aggregation as follows: 
\begin{equation}\label{eq:relation_aware}
    \mathbf{h}_{v}^{(l)}= \textnormal{ReLU}\left(\textnormal{AGG}_{all}^{(l)}\left(\mathbf{h}_{v}^{(l-1)}  \oplus \{p_{r}^{(l)} \cdot \mathbf{h}_{v, r}^{(l)} \}|_{r=1}^{R} \right)\right),
\end{equation}
where $\mathbf{h}_{v}^{(l-1)}$ is the center node embedding at the previous layer,
$\mathbf{h}_{v, r}^{(l)}$ is the intra-relation neighbor embedding at the $l$-th layer and $p_{r}^{(l)}$ is filtering threshold of relation $r$ which is directly used as its inter-relation aggregation weight.
$\oplus$ denotes the embedding summation operation. $\textnormal{AGG}^{l}_{all}$ can be any type of aggregator, and we test them in Section~\ref{sec05:overall_eval}.

\begin{algorithm}[b]
\caption{\textbf{CARE-GNN:} Camouflage Resistant GNN.}
\label{alg:care-gnn}
\SetKwData{False}{False}\SetKwData{This}{this}\SetKwData{Up}{up}
\SetKwFunction{Union}{Union}\SetKwFunction{FindCompress}{FindCompress}
\SetKwInOut{Input}{Input}\SetKwInOut{Output}{Output}

\Input{An undirected multi-relation graph with node features and labels: $\mathcal{G}=\left\{\mathcal{V}, \mathbf{X}, \{\mathcal{E}_{r}\}|_{r=1}^{R}, Y\right\}$;
\\ Number of layers, batches, epochs: $L, B, E$;
\\Parameterized similarity measures: $\{S^{(l)}(\cdot, \cdot)\}|_{l=1}^{L}$;
\\Filtering thresholds: $P=\{p^{(l)}_{1}, \dots, p^{(l)}_{R}\}|_{l=1}^{L}$;
\\Intra-R aggregators: $\{\textnormal{AGG}_{r}^{(l)}\}|_{r=1}^{R}, \forall l\in\{1,\dots,L\}$;
\\Inter-R aggregators: $\{\textnormal{AGG}_{all}^{(l)}\}, \forall l\in\{1,\dots,L\}$.}

\Output{Vector representations $\mathbf{z}_{v}, \forall v \in \mathcal{V}_{train}$.}

\BlankLine

\tcp{Initialization}

$\mathbf{h}_{v}^{0} \leftarrow \mathbf{x}_{v}, \forall v\in \mathcal{V};$ $p_{r}^{(0)}=0.5, \mathcal{E}_{r}^{(0)} = \mathcal{E},$ $\forall r\in\{1,\dots,R\}$\;

\For(\tcp*[f]{Train CARE-GNN}){$e = 1, \cdots, E$}{
    \For{$b = 1, \cdots, B$}{
        \For{$l = 1, \cdots, L$}{
            \For{$r = 1, \cdots, R$}{
        
                $S^{(l)}(v, v^{\prime}) \leftarrow$ Eq.~(\ref{eq:similarity}) , $\forall (v, v^{\prime})\in \mathcal{E}_{r}^{(l-1)}$\;
            
                $\mathcal{E}_{r}^{(l)} \leftarrow$ \textit{top-p} sampling (Section~\ref{sec04:sampling})\;

                $\mathbf{h}_{v,r}^{(l)} \leftarrow$ Eq.~(\ref{eq:aggregator}) $\forall v \in \mathcal{V}_{b}$; \tcp*[f]{Intra-R AGG}\ 
        
            }
        
            $\mathbf{h}_{v}^{(l)} \leftarrow$ Eq.~(\ref{eq:relation_aware}) $\forall v \in \mathcal{V}_{b}$; \tcp*[f]{Inter-R AGG}\ 
        
            $\mathcal{L}^{(1)}_{\textnormal{Simi}} \leftarrow$ Eq.~(\ref{eq:simi_loss}); \tcp*[f]{Simi loss}\ 
        }

        $\mathbf{z}_{v} \leftarrow \mathbf{h}_{v}^{(L)}, \forall v \in \mathcal{V}_{b}$; \tcp*[f]{Batch node embeddings}\
        
        $\mathcal{L}_{\textnormal{GNN}} \leftarrow$ Eq.~(\ref{eq:gnn_loss}); \tcp*[f]{GNN loss}\ 
    
        $\mathcal{L}_{\textnormal{CARE}} \leftarrow$ Eq.~(\ref{eq:total_loss}); \tcp*[f]{CARE-GNN loss}\ 
    }
    
    \For(\tcp*[f]{RL Module}){$l = 1, \cdots, L$}{ 
        \For{$r = 1, \cdots, R$}{
            \If{\textnormal{Eq.~(\ref{eq:terminal}) is} \False}{
                
                $f(p_{r}^{(l)}, a_{r}^{(l)})^{(e)} \leftarrow$ Eqs.(\ref{eq:avg_distance}) and (\ref{eq:threshold_rewards})\;
                $p_{r}^{(l)} \leftarrow p_{r}^{(l)} + f(p_{r}^{(l)}, a_{r}^{(l)})^{(e)} \cdot \tau$\;
            }
        }
    }
}

\end{algorithm}

\subsection{Proposed CARE-GNN}
\label{sec04:optimization}

\noindent \textbf{Optimization.} For each node $v$, its final embedding is the output of the GNN at the last layer $\mathbf{z}_{v} = \mathbf{h}_{v}^{(L)}$.
We can define the loss of GNN as a cross-entropy loss function:
\begin{equation}\label{eq:gnn_loss}
    \mathcal{L}_{\textnormal{GNN}} = \sum_{v\in \mathcal{V}}-\log\left(y_{v}\cdot\sigma(MLP(\mathbf{z}_{v}))\right).
\end{equation}
Together with the loss function of the similarity measure in Eq.~(\ref{eq:simi_loss}), we define the loss of CARE-GNN as:
\begin{equation}\label{eq:total_loss}
    \mathcal{L}_{\textnormal{CARE}} = \mathcal{L}_{\textnormal{GNN}} + \lambda_{1}\mathcal{L}^{(1)}_{\textnormal{Simi}} + \lambda_{2}||\Theta||_{2},
\end{equation}
where $||\Theta||_{2}$ is the $L2$-norm of all model parameters, $\lambda_{1}$ and $\lambda_{2}$ are weighting parameters.
Since the neighbor filtering process at the first layer is critical to both GNN and similarity measures in the following layers, we only use the similarity measure loss at the first layer to update the parameterized similarity measure in Eq. (\ref{eq:similarity}).

\vspace{1mm}
\noindent \textbf{Algorithm Description.} Algorithm~\ref{alg:care-gnn} shows the training process of the proposed CARE-GNN.
Given a multi-relational fraud graph, we employ the mini-batch training technique~\cite{goyal2017accurate} as the result of its large scale.
In the beginning, we randomly initialize the parameters of the similarity measure module and GNN module. We initialize all filtering thresholds as $0.5$ (Line 2).
For each batch of nodes, we first compute the neighbor similarities using Eq.~(\ref{eq:similarity}) (Line 7) and then filter the neighbors using \textit{top-p} sampling (Line 8).
Then, we can compute the intra-relation embeddings (Line 9), inter-relation embeddings (Line 10), loss functions (Lines 11-14) for the current batch, respectively.
As for the RL process, we assign random actions for the first epoch since it has no reference.
From the second epoch, we update $p_{r}^{(l)}$ according to Lines 15-19.

% \vspace{1mm}
% \noindent \textbf{Discussion.} \textcolor{red}{layers and time complexity.}

\section{Experiments}
\label{sec05:exp}
% \textcolor{red}{Highlight the findings.}
In the experiment section, we mainly present:
\begin{itemize}%[leftmargin=*]
    \item how we construct multi-relation graphs upon different fraud data (Section~\ref{sec05:graph_build});
    
    \item camouflage evidences in real-world fraud data (Section~\ref{sec05:camoflage});
    
    \item the performance comparison over baselines and CARE-GNN variants (Section~\ref{sec05:overall_eval});
    
    \item the learning process and explanation of the RL algorithm (Section~\ref{sec05:training_analysis});
    
    \item the sensitivity study of hyper-parameters and their effects on model designing (Section~\ref{sec05:sensitivity}).
\end{itemize}

\subsection{Experimental Setup}
\label{sec05:setup}

% \begin{table*}[h]
% \centering
% \caption{The statistics of datasets.}
% \resizebox{0.9\linewidth}{!}{%
% \begin{tabular}{c|cccccccc}  
% \hline
% \textbf{Dataset}  & \textbf{\%Fraud} & \textbf{\#Nodes} & \begin{tabular}[c]{@{}c@{}} \textbf{Node Feature} \\ \textbf{Dimension}\end{tabular} &\textbf{Relation} &  \textbf{\#Edges} &  \begin{tabular}[c]{@{}c@{}}\textbf{Avg. Node} \\ \textbf{Degree}\end{tabular} & \begin{tabular}[c]{@{}c@{}} \textbf{Avg. Feature} \\ \textbf{Similarity}\end{tabular}  & \begin{tabular}[c]{@{}c@{}} \textbf{Avg. Label} \\ \textbf{Similarity}\end{tabular}  \\ 
% \hline
%  \multirow{4}{*}{\begin{tabular}[c]{@{}c} Yelp \end{tabular}} &  \multirow{4}{*}{\begin{tabular}[c]{@{}c} \%14.5 \end{tabular}}& \multirow{4}{*}{\begin{tabular}[c]{@{}c} 45,954 \end{tabular}} & \multirow{4}{*}{\begin{tabular}[c]{@{}c} 32 \end{tabular}}& Review-User-Review (\textit{R-U-R}) &  98,630  & 2 & 0.90 & 0.91 \\
% & & & & Review-Time-Review (\textit{R-T-R}) & 1,147,232  & 25 &0.89 & 0.05\\
% & & &&Review-Star-Review (\textit{R-S-R}) & 6,805,486  &148& 0.89 & 0.05\\
% &&  & &\textit{ALL} & 7,693,958 & 167 & 0.89& 0.07\\
% \hline
% \end{tabular}}
% \label{tab:stat}
% \end{table*}

\begin{table}[h]
\small
\centering
\caption{Dataset and graph statistics.}
\resizebox{0.98\linewidth}{!}{%
\begin{tabular}{c|lcccc}  
\hline
\multicolumn{2}{r}{\begin{tabular}[c]{@{}c@{}}\textbf{\#Nodes} \\ (\textbf{Fraud\%})\end{tabular}} & \textbf{Relation}  & \textbf{\#Edges}  & \begin{tabular}[c]{@{}c@{}} \textbf{Avg. Feature} \\ \textbf{Similarity}\end{tabular}  & \begin{tabular}[c]{@{}c@{}} \textbf{Avg. Label} \\ \textbf{Similarity}\end{tabular}  \\ 

\hline
 \multirow{4}{*}{\rotatebox[origin=c]{90}{\textbf{Yelp}}} & \multirow{4}{*}{\begin{tabular}[c]{@{}c@{}} 45,954 \\ (14.5\%)\end{tabular}}  & \textit{R-U-R} & 49,315   & 0.83 & 0.90 \\
&  & \textit{R-T-R} & 573,616   &0.79 & 0.05\\
&  & \textit{R-S-R} & 3,402,743  & 0.77 & 0.05\\
&  & \textit{ALL} & 3,846,979   & 0.77& 0.07\\
\hline
\multirow{4}{*}{\rotatebox[origin=c]{90}{\textbf{Amazon}}} & \multirow{4}{*}{\begin{tabular}[c]{@{}c@{}} 11,944 \\ (9.5\%)\end{tabular}}  & \textit{U-P-U} & 175,608   & 0.61 & 0.19 \\
&  & \textit{U-S-U} & 3,566,479   &0.64 & 0.04\\
&  & \textit{U-V-U} & 1,036,737  & 0.71 & 0.03\\
&  & \textit{ALL} & 4,398,392   & 0.65 & 0.05\\
\hline
\end{tabular}}
\label{tab:stat}
\end{table}

\subsubsection{Dataset.}
We use the Yelp review dataset~\cite{Rayana2015} and Amazon review dataset~\cite{mcauley2013amateurs} to study the fraudster camouflage and GNN-based fraud detection problem.
The Yelp dataset includes hotel and restaurant reviews filtered (spam) and recommended (legitimate) by Yelp.
The Amazon dataset includes product reviews under the Musical Instruments category.
Similar to~\cite{zhang2020gcn}, we label users with more than 80\% helpful votes as benign entities and users with less than 20\% helpful votes as fraudulent entities.
Though previous works have proposed other fraud datasets like Epinions~\cite{kumar2018rev2} and Bitcoin~\cite{weber2019anti}, they only contain graph structures and compacted features, with which we cannot build meaningful multi-relation graphs.
% 1) it includes $user\_id$, $product\_id$, $review\;text$, $rating$ and $timestamp$ which are sufficient to construct graph with multiple relations. Meanwhile, those attributes are common in many fraud detection tasks;
% 2) previous studies~\cite{Rayana2015,Mukherjee:2013uk} have proposed many useful features based on the semantic and behavior information on Yelp dataset, which can be directly leveraged as node features;
% 3) recent works~\cite{wang2019fdgars, li2019spam} have employed GNN to spot spam reviews on other platforms, which shows the feasibility of applying GNN to spam detection task.
In this paper, we conduct a spam review detection (fraudulent user detection resp.) task on the Yelp dataset (Amazon dataset resp.), which is a binary classification task.
We take 32 handcrafted features from~\cite{Rayana2015} (25 handcrafted features from~\cite{zhang2020gcn} resp.) as the raw node features for Yelp (Amazon resp.) dataset.
Table~\ref{tab:stat} shows the dataset statistics.

\subsubsection{Graph Construction}
\label{sec05:graph_build}
\noindent \textbf{Yelp:} based on previous studies~\cite{Rayana2015, Mukherjee:2013uk} which show that opinion fraudsters have connections in user, product, review text, and time,
we take reviews as nodes in the graph and design three relations: 1) \textit{R-U-R}: it connects reviews posted by the same user; 2) \textit{R-S-R}: it connects reviews under the same product with the same star rating (1-5 stars); 3) \textit{R-T-R}: it connects two reviews under the same product posted in the same month.
\textbf{Amazon:} similarly, we take users as nodes in the graph and design three relations: 1) \textit{U-P-U}: it connects users reviewing at least one same product; 2) \textit{U-S-V}: it connects users having at least one same star rating within one week; 3) \textit{U-V-U}: it connects users with top 5\% mutual review text similarities (measured by TF-IDF) among all users.
The number of edges belonging to each relation is shown in Table~\ref{tab:stat}.

\subsubsection{Baselines.}
To verify the ability of CARE-GNN in alleviating the negative influence induced by camouflaged fraudsters, we compare it with various GNN baselines under the semi-supervised learning setting. We select GCN~\cite{kipf2016semi}, GAT~\cite{velivckovic2017graph}, RGCN~\cite{schlichtkrull2018modeling}, and GraphSAGE~\cite{hamilton2017inductive} to represent general GNN models. We choose GeniePath~\cite{liu2019geniepath}, Player2Vec~\cite{zhang2019key}, SemiGNN~\cite{wang2019semi}, and GraphConsis~\cite{liu2020alleviating} as four state-of-the-art GNN-based fraud detectors. Their detailed introduction can be found in Section~\ref{sec02:RW}. We also implement several variants of CARE-GNN: CARE-\textit{Att}, CARE-\textit{Weight}, and CARE-\textit{Mean}, and they differ from each other in Attention~\cite{velivckovic2017graph}, Weight~\cite{liu2018heterogeneous}, and Mean~\cite{hamilton2017inductive} inter-relation aggregator respectively.

Among those baselines, GCN, GAT, GraphSAGE, and GeniePath are run on homogeneous graphs (i.e., Relation \textit{ALL} in Table~\ref{tab:stat}) where all relations are merged together. Other models are run on multi-relation graphs where they handle information from different relations in their approaches.

\begin{table*}[h]
\small
\centering
\caption{Fraud detection performance (\%) on two datasets under different percentage of training data.}
\resizebox{0.95\linewidth}{!}{%
\begin{tabular}{|c|c|c|c|c|c|c|c|c|c|c||c|c|c|c|}  
\hline
& \textbf{Metric}  & \textbf{Train\%}  & \textbf{GCN} & \textbf{GAT} & \textbf{RGCN} & \begin{tabular}[c]{@{}c@{}} \textbf{Graph-}\\ \textbf{SAGE} \end{tabular}   & \begin{tabular}[c]{@{}c@{}} \textbf{Genie-}\\ \textbf{Path} \end{tabular} & \begin{tabular}[c]{@{}c@{}} \textbf{Player-}\\ \textbf{2Vec} \end{tabular} & \begin{tabular}[c]{@{}c@{}} \textbf{Semi-}\\ \textbf{GNN} \end{tabular} & \begin{tabular}[c]{@{}c@{}} \textbf{Graph-}\\ \textbf{Consis} \end{tabular} & \begin{tabular}[c]{@{}c@{}} \textbf{CARE-}\\ \textbf{\textit{Att}} \end{tabular} &  \begin{tabular}[c]{@{}c@{}} \textbf{CARE-}\\ \textbf{\textit{Weight}} \end{tabular} &  \begin{tabular}[c]{@{}c@{}} \textbf{CARE-}\\ \textbf{\textit{Mean}} \end{tabular} &  \begin{tabular}[c]{@{}c@{}} \textbf{CARE-}\\ \textbf{GNN} \end{tabular} \\ 

\hline
  \multirow{8}{*}{\rotatebox[origin=c]{90}{\textbf{Yelp}}} & \multirow{4}{*}{AUC} & 5\%  & 54.98  & 56.23 & 50.21  & 53.82   & 56.33   & 51.03  & 53.73  & 61.58   & 66.08   & 71.10  & 69.83 & \textbf{71.26}\\
  & & 10\%  & 50.94 & 55.45   & 55.12  & 54.20   & 56.29   & 50.15  & 51.68  & 62.07   & 70.21   & 71.02  & 71.85 & \textbf{73.31}\\
  & & 20\%  & 53.15 & 57.69   & 55.05  & 56.12  & 57.32   & 51.56  & 51.55  & 62.31  & 73.26   & 74.32  & 73.32 & \textbf{74.45}\\
  & & 40\%  & 52.47 & 56.24   & 53.38  & 54.00   & 55.91   & 53.65  & 51.58  & 62.07   & 74.98   & 74.42  & 74.77 & \textbf{75.70}\\
\cline{2-15}
  & \multirow{4}{*}{Recall}& 5\%  & 53.12 & 54.68   & 50.38  & 54.25   & 52.33   & 50.00  & 52.28  & 62.60   & 63.52   & 66.64  & \textbf{68.09} & 67.53\\
  & & 10\%  & 51.10 & 52.34   & 51.75  & 52.23   & 54.35   & 50.00  & 52.57  & 62.08   & 67.38   & 68.35  & \textbf{68.92} & 67.77\\
  & & 20\%  & 53.87 & 53.20   & 50.92  & 52.69   & 54.84   & 50.00  & 52.16  & 62.35   & 68.34   & 69.07  & \textbf{69.48} & 68.60\\
  & & 40\%  & 50.81 & 54.52   & 50.43  & 52.86   & 50.94   & 50.00  & 50.59  & 62.08   & 71.13   & 70.22  & 69.25 & \textbf{71.92}\\
\hline
\hline
  \multirow{8}{*}{\rotatebox[origin=c]{90}{\textbf{Amazon}}} & \multirow{4}{*}{AUC} & 5\%  & 74.44 & 73.89   & 75.12  & 70.71   & 71.56   & 76.86  & 70.25  & 85.46   & 89.49  & 89.36 & 89.35 & \textbf{89.54}\\
  & & 10\%  & 75.25 & 74.55   & 74.13  & 73.97   & 72.23   & 75.73  & 76.21  & 85.29    & \textbf{89.58}  & 89.37 & 89.43 & 89.44\\
  & & 20\%  & 75.13 & 72.10   & 75.58  & 73.97   & 71.89   & 74.55  & 73.98  & 85.50    & 89.58  & \textbf{89.68} & 89.34 & 89.45\\
  & & 40\%  & 74.34  & 75.16 & 74.68 & 75.27   & 72.65  & 56.94 & 70.35 & 85.50   & 89.70  & 89.69 & 89.52 & \textbf{89.73}\\
\cline{2-15}
  & \multirow{4}{*}{Recall}& 5\%  & 65.54 & 63.22   & 64.23  & 69.09   & 65.56   & 50.00  & 63.29  & 85.49   & 88.22  & 88.31 & 88.02 & \textbf{88.34}\\
  & & 10\%  & 67.81 & 65.84   & 67.22  & 69.36   & 66.63   & 50.00  & 63.32  & 85.38   & 87.87  & \textbf{88.36} & 88.12 & 88.29\\
  & & 20\%  & 66.15 & 67.13   & 65.08  & 70.30   & 65.08   & 50.00  & 61.28  & 85.59   & 88.40  & \textbf{88.60} & 88.00 & 88.27\\
  & & 40\%  & 67.45   & 65.51 & 67.68 & 70.16   & 65.41  & 50.00 & 62.89 & 85.53   & 88.41  & 88.45 & 88.22 & \textbf{88.48}\\
  \hline
\end{tabular}}
\label{tab:overall}
\end{table*}

\subsubsection{Experimental Setting}
\label{sec05:experimental_setting}
From Table~\ref{tab:stat}, we can see that the percentage of fraudsters are small in both datasets.
Meanwhile, real-world graphs usually have great scales.
To improve the training efficiency and avoid overfitting, we employ mini-batch training~\cite{goyal2017accurate} and under-sampling~\cite{liu2008exploratory} techniques to train CARE-GNN and other baselines.
Specifically, under each mini-batch, we randomly sample the same number of negative instances as the number of positive instances.
We also study the sample ratio sensitivity in Section~\ref{sec05:sensitivity}.

We use unified node embedding size (64), batch size (1024 for Yelp, 256 for Amazon), number of layers(1), learning rate (0.01), optimizer (Adam), and L2 regularization weight ($\lambda_2=0.001$) for all models.
For CARE-GNN and its variants, we set the RL action step size ($\tau$) as 0.02 and the similarity loss weight ($\lambda_1$) as 2.
In Section~\ref{sec05:sensitivity}, we present the sensitivity study for the number of layers, embedding size, and $\lambda_1$.   

\subsubsection{Implementation}
For the GCN, GAT, RGCN, GraphSAGE, GeniePath, we use the source code provided by authors.
For Player2Vec, SemiGNN, and GraphConsis, we use the open-source implementations\footnote{\url{https://github.com/safe-graph/DGFraud}}.
We implement CARE-GNN with Pytorch. All models are running on Python 3.7.3, 2 NVIDIA GTX 1080 Ti GPUs, 64GB RAM, 3.50GHz Intel Core i5 Linux desktop.

\subsubsection{Evaluation Metric}
Since the Yelp dataset has imbalanced classes, and we focus more on fraudsters (positive instances),
like previous work~\cite{Rayana2015}, we utilize ROC-AUC (AUC) and Recall to evaluate the overall performance of all classifiers.
AUC is computed based on the relative ranking of prediction probabilities of all instances, which could eliminate the influence of imbalanced classes.

\subsection{Camouflage Evidence}
\label{sec05:camoflage}
We analyze fraudster camouflage using two metrics introduced in~\cite{liu2020alleviating}. 
For the feature camouflage, we compute the feature similarity of neighboring nodes based on their feature vectors' Euclidean distance, ranging from $0$ to $1$.
The average feature similarity is normalized w.r.t. the total number of edges, which is presented in Table~\ref{tab:stat}.
We observe that the averaged similarity scores under all relations are high.
High feature similarity implies that fraudsters camouflage their features in a similar way to benign nodes.
Moreover, the minor feature similarity difference across different relations proves that the unsupervised similarity measure cannot effectively discriminate fraudsters and benign entities.
For instance, the label similarity difference between \textit{R-U-R} and \textit{R-T-R} is 0.85, but the feature similarity difference is only 0.04.  

For the relation camouflage, we study it by calculating the label similarity based on whether two connected nodes have the same label.
The label similarity is normalized w.r.t. the total number of edges.
The average label similarity for each relation is shown in Table~\ref{tab:stat}.
High label similarity score implies that the fraudsters fail to camouflage, and low score implies that fraudsters camouflage successfully.
We observe that only \textit{R-U-R} relation has a high label similarity score, while the other relations have label similarity scores less than 20\%.
It suggests that we need to select different amounts of neighbors for different relations to facilitate the GNN aggregation process.
Meanwhile, we should distinguish relations in order to prevent fraudsters from camouflaging.

\subsection{Overall Evaluation}
\label{sec05:overall_eval}

Table~\ref{tab:overall} shows the performance of proposed CARE-GNN and various GNN baselines under the fraud detection task on two datasets.
We report the best testing results after thirty epochs. 
We observe that CARE-GNN outperforms other baselines under most of the training proportions and metrics.

\vspace{1mm}
\noindent \textbf{Single-relation vs. Multi-relation.}
Among all GNN baselines in Table~\ref{tab:overall}, GCN, GAT, GraphSAGE, and GeniePath run on single-relation (i.e., homogeneous) graph where all relations are merged together (\textit{ALL} in Table~\ref{tab:stat}).
Other baselines are built upon multi-relation graphs.
The performances of single-relation GNNs are better than Player2Vec and SemiGNN, which indicates previously designed fraud detection methods are not suitable for multi-relation graphs.
Among the multi-relation GNNs, GraphConsis outperforms all other multi-relation GNNs.
The reason is that GraphConsis samples the neighbors based on the node features before aggregating them. Better than GraphConsis, CARE-GNN and its variants adopt parameterized similarity measure and adaptive sampling thresholds, which could better identify and filter camouflaged fraudsters. 
It demonstrates that neighbor filtering is critical to GNNs when the graph contains many noises (i.e., dissimilar/camouflaged neighbors).
Also, CARE-GNN has higher scores than all single-relation GNNs, suggesting that a noisy graph undermines the performance of multi-relation GNNs.
A possible reason is the higher complexity of multi-relation GNNs comparing to single-relation ones.

\begin{figure*}
     \centering
     \begin{subfigure}[b]{0.98\textwidth}
         \centering
         \includegraphics[width=\textwidth]{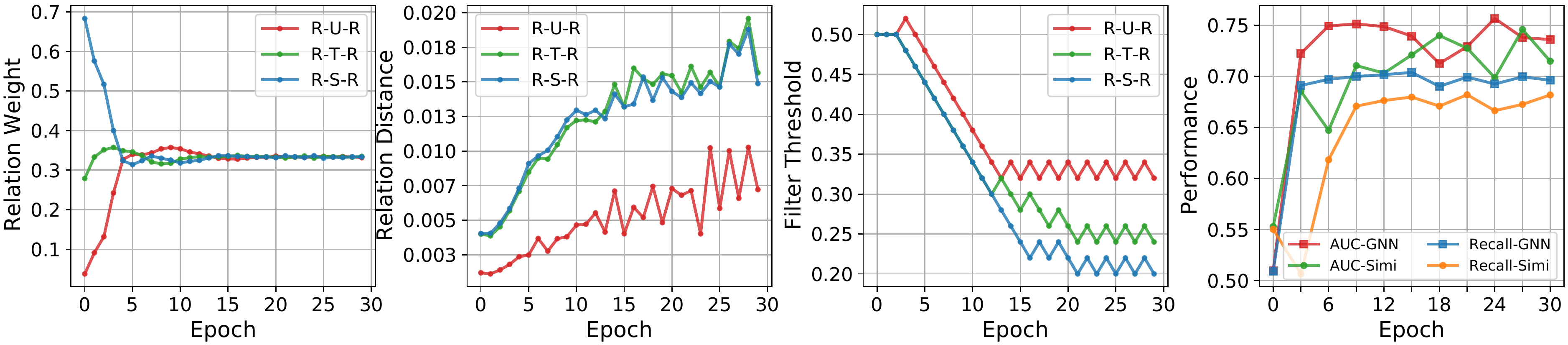}
         \label{fig:yelp_rl}
     \end{subfigure}
    \bigskip
     \begin{subfigure}[b]{0.98\textwidth}
         \centering
         \includegraphics[width=\textwidth]{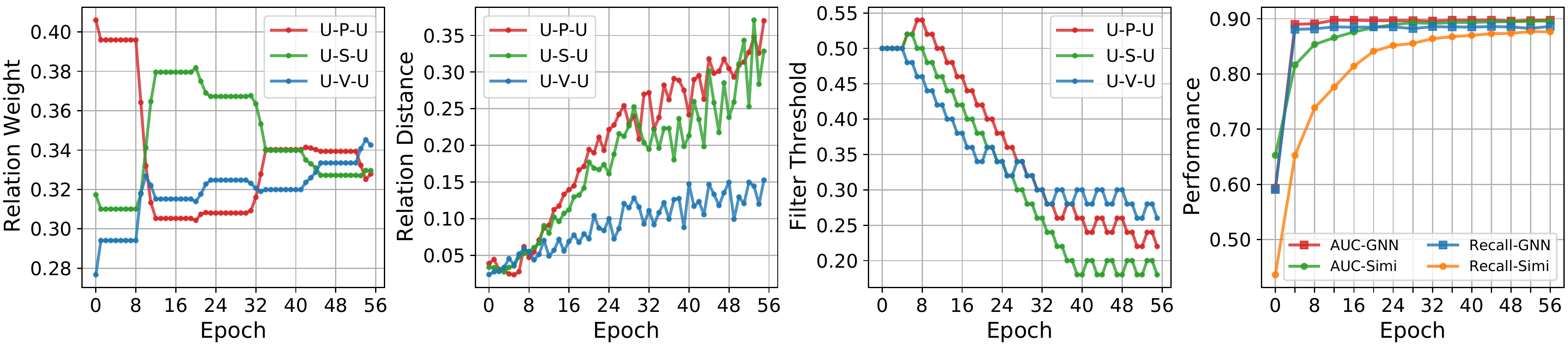}
         \label{fig:amz_rl}
     \end{subfigure}
     \vspace*{-5mm}
        \caption{The training process and testing performance of CARE-\textit{Weight} on Yelp (upper) and Amazon (lower) dataset.} 
        \label{fig:training}
\end{figure*}

\vspace{1mm}
\noindent \textbf{Training Percentage.}
From Table~\ref{tab:overall}, there is little performance gain for GNNs when increasing the training percentages.
It demonstrates the advantage of semi-supervised learning, where a small amount of supervised signals is enough to train a good model.
Meanwhile, with informative handcraft features as inputs for two datasets, GNNs are much easier to learn high-quality embeddings. 
% For the baselines, we could see that the performance of most models become worse under the 40\% training data which is counter-intuitive.
% Recall the camouflage evidence we introduced in Section~\ref{sec05:camoflage} and Table~\ref{tab:stat}, if we do not select the similar neighbors under \textit{R-T-R} and \textit{R-S-R},
% increasing the training data amount will provide more misleading signals which exacerbate those baselines towards classifying fraudsters as begin ones.

\vspace{1mm}
\noindent \textbf{CARE-GNN Variants.}
The last four columns of Table~\ref{tab:overall} show the performance of CARE-GNN and its variants with different inter-relation aggregators.
It is observed that those four models have similar performances under most training percentages and metrics.
It verifies our assumption in Section~\ref{sec04:aggregation} that the attention coefficients and relation weights will become unnecessary when we select similar neighbors under all relations.
Moreover, for the Yelp dataset, the CARE-\textit{Att} has worse performances under a smaller training percentage (e.g., 5\%).
While for CARE-GNN, since it does not need to train extra attention weights, it attains the best performance against other variants. 
The first column of Figure~\ref{fig:training} presents more evidence that the relation weights finally become equal for all relations under both datasets.
The better performance of CARE-GNN comparing to CARE-\textit{Mean} shows that keeping the filtering threshold as inter-relation aggregation weights could enhance the GNN performance and reduce model complexity.

\vspace{1mm}
\noindent \textbf{GNN vs. Similarity Measure (Figure~\ref{fig:training} Column 4).}
Figure~\ref{fig:training} Column 4 shows the testing performances solely based on the outputs of the GNN module and similarity measure module during training.
For the Yelp dataset, GNN has better AUC and Recall than the similarity measure, which suggests that leveraging the structural information benefits the model to classify fraud and benign entities. 
For Amazon, the performance of GNN and the similarity measure are comparable with each other.
It is because the input features provide enough information to discriminate fraudsters.

\subsection{RL Process Analysis}
\label{sec05:training_analysis}

In this paper, we jointly train the similarity measure and GNN together and employ RL to find the neighbor filtering thresholds adaptively.
To present the RL process from different perspectives, in Figure~\ref{fig:training}, we plot the updating process of three parameters without terminating the RL process during training CARE-\textit{Weight}.
Since CARE-\textit{Weight} learns the aggregation weight for each relation, plotting its training process instead of CARE-GNN could help understand the effects of our proposed GNN enhancement modules.
During training, we also test the model every three epochs for Yelp (four epochs for Amazon) and plot the testing performance for both GNN and similarity measure at the last column of Figure~\ref{fig:training}.

\vspace{1mm}
\noindent \textbf{Relation Weights (Figure~\ref{fig:training} Column 1).}
We observe that the randomly initialized relation aggregation weights gradually converge to the same value as the neighbor selector updates its filtering thresholds and selects more similar neighbors under each relation.
When neighbors under each relation provide similar information, their aggregation weights will be similar as well.

\vspace{1mm}
\noindent \textbf{Relation Distance (Figure~\ref{fig:training} Column 2).}
As the training epoch increases, it is clearly that the differences between neighbor distances under each relation (computed by Eq. (\ref{eq:avg_distance})) become larger and comparable to each other.
The reason is that the GNN projects the node embeddings to a broader range of space and makes them more distinguishable.
As the model filters more noisy neighbors, the average distance across different relations become closer.

\vspace{1mm}
\noindent \textbf{Neighbor Filtering Threshold (Figure~\ref{fig:training} Column 3).}
We take 0.02 as the action step size; all thresholds are updated and converge to different values.
When the filtering threshold oscillates for several rounds, it reaches the terminal condition in Eq. (\ref{eq:terminal}).
For different datasets, the proposed RL algorithm could adaptively find the optimal filtering thresholds.

\begin{figure}
     \centering
         \includegraphics[width=0.485\textwidth]{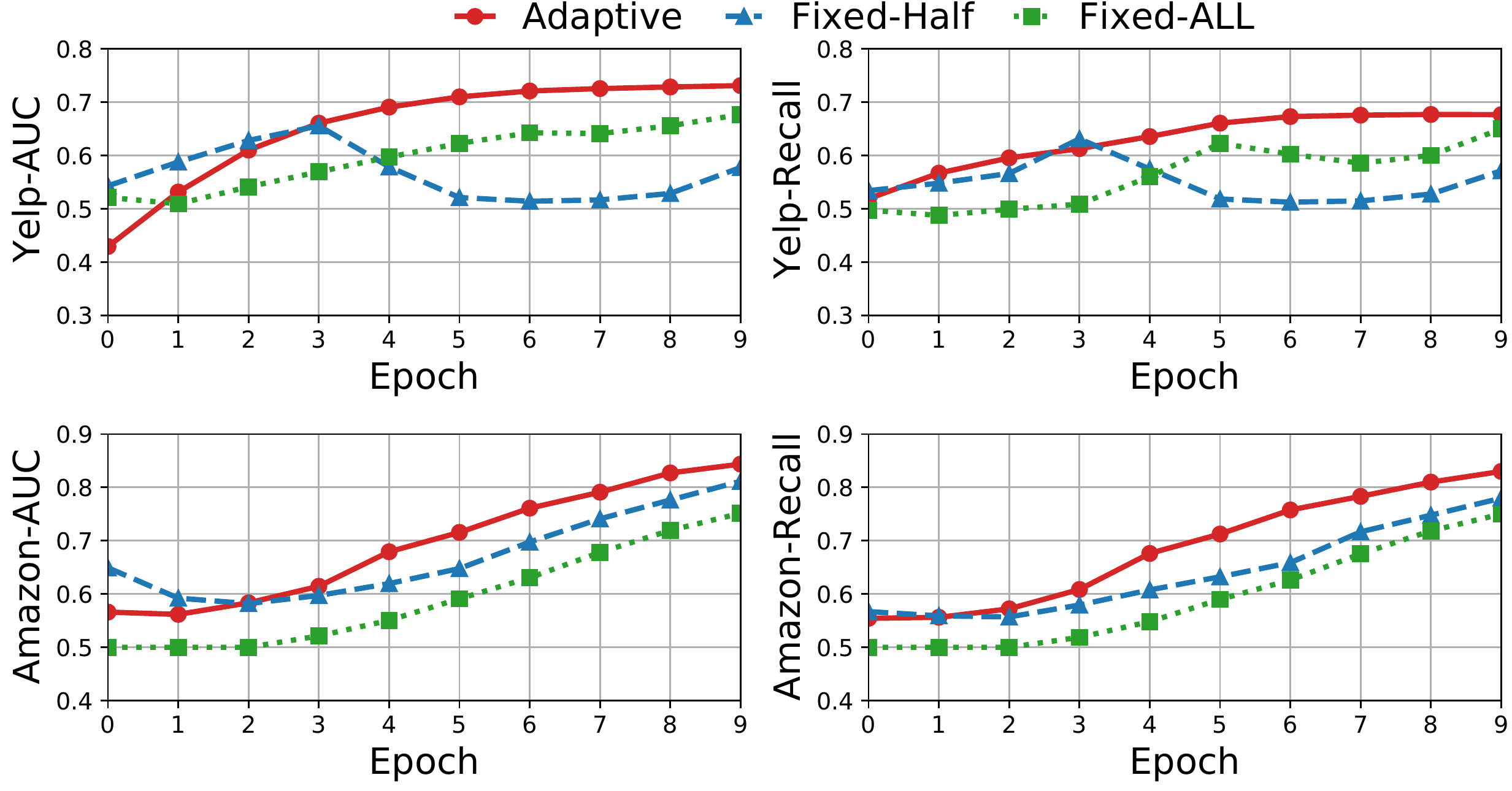}
         \caption{The testing AUC and Recall for CARE-GNN with different neighbor filtering methods during training.}
        \label{fig:adaptive}
\end{figure}

To demonstrate the advantage of the optimal neighbor filtering thresholds found by RL, in Figure~\ref{fig:adaptive}, we plot the testing performances of three different neighbor selection criteria under two datasets. 
$\texttt{Adaptive}$ filters neighbors using converged thresholds found by RL (as shown in Figure~\ref{fig:training} Column 3); $\texttt{Fixed-Half}$ keeps the \textit{top} 50\% similar neighbors under each relation and $\texttt{Fixed-All}$ keeps all neighbors without filtering.   
It is illustrated that CARE-GNN with adaptive filtering thresholds is optimized faster than the other two neighbor selectors.
Meanwhile, it has a better and smoother performance during training. 
It verifies the effectiveness of the proposed RL algorithm, which is able to find informative neighbors under each relation.

\subsection{Hyper-parameter Sensitivity}
\label{sec05:sensitivity}

\begin{figure}
     \centering
         \includegraphics[width=0.47\textwidth]{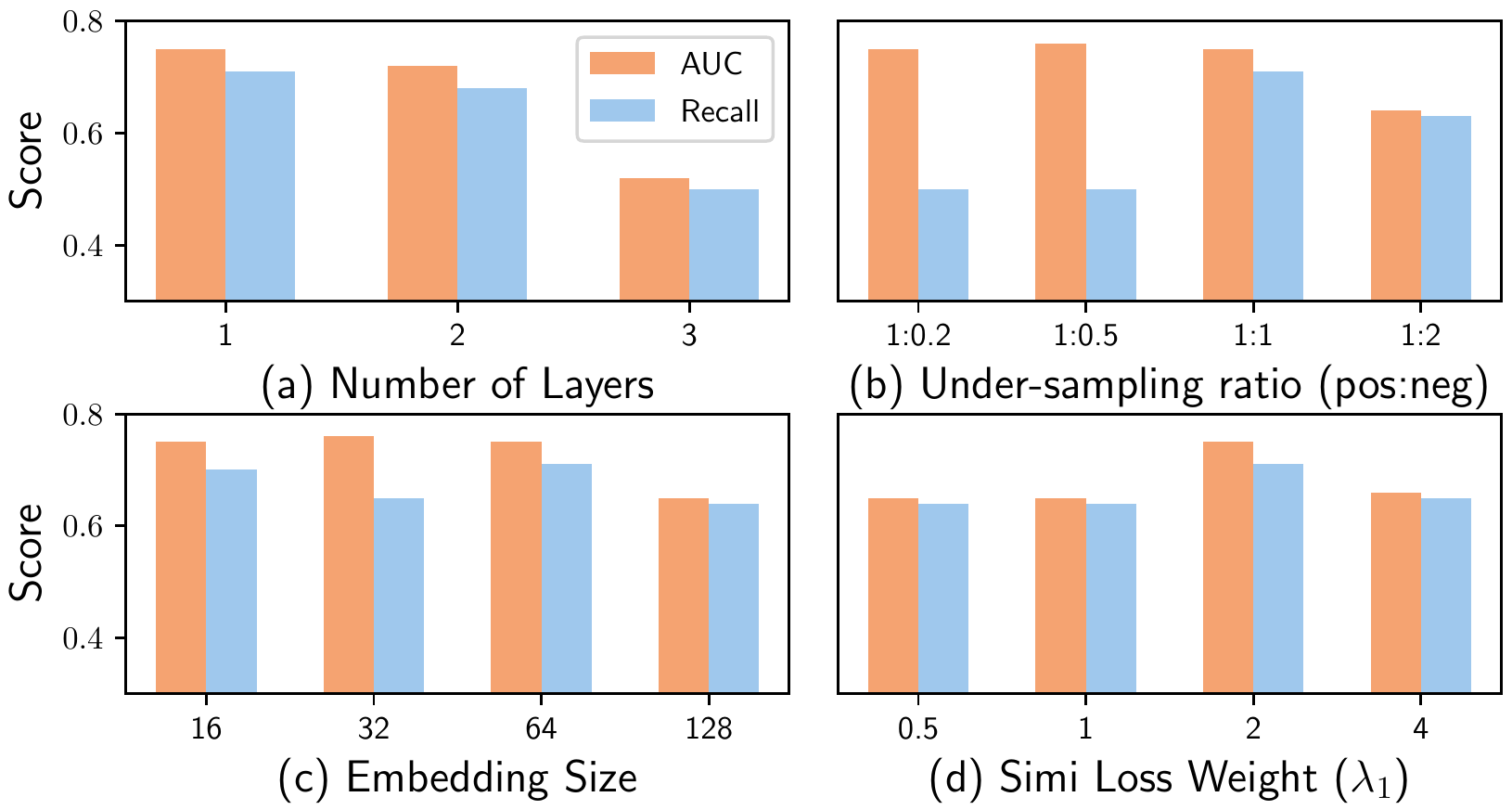}
         \caption{Parameter Sensitivity. For each parameter configuration, only the best results among 30 epochs are recorded.}
        \label{fig:sensitivity}
\end{figure}

% AUC-recall bar chart The number of layers fixed thresholds, 
Figure~\ref{fig:sensitivity} shows the testing performance of CARE-GNN regarding four hyper-parameters on the Yelp dataset.
From Figure~\ref{fig:sensitivity}(a), we observe that increasing the number of layers barely improves the performance of CARE-GNN.
For the three-layer model, the CARE-GNN suffers the overfitting problem (Recall = 0.5).
Therefore, the one-layer model is not only able to save the computational cost but also achieve better classification results.
Figure~\ref{fig:sensitivity}(b) presents the CARE-GNN performance under different under-sampling ratios as introduced in Section~\ref{sec05:experimental_setting}.
Note that CARE-GNN is tested on an imbalanced test set.
Moreover, CARE-GNN is overfitted when negative instances are less than positive ones (under 1:0.2 and 1:0.5, Recalls are equal to 0.5).
An equal under-sampling ratio guarantees a good and fair performance of CARE-GNN.
Figure~\ref{fig:sensitivity}(c) shows the influence of different embedding sizes. Embedding sizes with 16, 32, and 64 have comparable performance.
Figure~\ref{fig:sensitivity}(d) illustrates the effects of different weighting values for the similarity loss ($\lambda_1$ in Eq. (\ref{eq:total_loss})).
When the weight of similarity loss is doubled compared to which of GNN loss, CARE-GNN reaches the best performance.
Therefore, the similarity measure is crucial for GNN training.

\subsection{Discussion}
\label{sec05:discussion}

Since the multi-relation graphs used in the experiments are very dense (average node degree $>150$), one-layer CARE-GNN (which aggregates one-hop neighbors) has already utilized abundant information and thus can achieve excellent performance.
CARE-GNN with more layers is suitable for sparse graphs.
We improve the computational efficiency using multiple approaches: the light-weight similarity measure, the classic and fast RL framework, positive-node based neighbor selector, no attention mechanism, and mini-batch training with under-sampling.
For CARE-GNN, each epoch only takes 17 seconds on Yelp (3 seconds on Amazon), and it has a great performance gain comparing to other baselines.

\section{Related Work}
\label{sec02:RW}

% Our study focuses on enhancing GNNs to be resistant to the camouflaged fraudsters. The related works can be categorized into two following topics: one is GNN and its enhancement, and another is GNN-based fraud detection.

% \subsection{GNN and Its Enhancement}
\noindent \textbf{GNN and Its Enhancement.} As the most popular deep learning framework on graph data, GNNs have two major types~\cite{wu2020comprehensive}: 1) Spectral-based GNNs (GCN~\cite{kipf2016semi}, AGCN~\cite{li2018adaptive}):
they turn a graph into a Laplacian matrix and make convolutional operations in the spectral domain.
2) Spatial-based GNNs (GAT~\cite{velivckovic2017graph}, GraphSAGE~\cite{ hamilton2017inductive}):
they propagate the information based on the spatial relation (i.e., the adjacent nodes).
Since spatial-based GNNs are more flexible, many GNN variants belong to this type.
The proposed CARE-GNN is a spatial-based GNN as well.
% state the shortcoming of attention and pooling

To enhance the GNN performance on graphs with noisy nodes. One approach is the \textit{graph structure learning} (GSL)~\cite{li2018adaptive, franceschi2019learning, chen2019deep}.
Those works learn new graph structures from original graphs, which could better render the latent connections between nodes.
% They conduct convolutional operations on the new graph.
Comparing to our work, those papers only investigate the single-relation benchmark datasets without camouflaged fraudsters.
Our model filters dissimilar neighbors instead of learning new structures.

Another approach is the \textit{metric learning}~\cite{chen2019label, hou2020measure}. Those works devise new metrics to measure the similarity between connect nodes and aggregate neighbors according to the metrics.
Among those works, \cite{chen2019label} proposes a neural network to predict the labels of neighboring nodes.
\cite{hou2020measure} devises two metrics to measure the average neighborhood similarity and label similarity in a graph.
However, those methods either have weak similarity metrics or fixed neighbor filtering thresholds, which need to be calibrated empirically.
CARE-GNN proposed by us is more flexible which could learn the similarity metric based on domain knowledge.
The relation filtering thresholds of CARE-GNN are optimized during training GNN which retains the end-to-end learning fashion.

GNN sampling methods~\cite{chen2018fastgcn, zou2019layer, zeng2019graphsaint} also filter the neighbors.
% Specifically, they leverage the importance sampling and variance reduction techniques to sample the nodes.
While these works only consider selecting representative nodes to accelerate GNN training.
For our work, taking account of domain knowledge and relational information, our goal is to filter dissimilar neighbors before aggregation, which could alleviate the negative effect of camouflaged fraudsters.

% SIGIR paper

% \subsection{GNN-based Fraud Detection}

%As an emerging research direction, there are only few previous works~\cite{liu2018heterogeneous, hu2019cash, wang2019fdgars, li2019spam, zhang2019key, wang2019semi} applying GNNs in fraud detection problems and most of them are used in opinion fraud~\cite{wang2019fdgars, li2019spam} and financial fraud~\cite{liu2018heterogeneous, hu2019cash, zhang2019key, wang2019semi}. 

\noindent \textbf{GNN-based Fraud Detection.} Many GNN-based fraud detectors transfer the heterogeneous data into homogeneous data before applying GNNs.
Fdgars~\cite{wang2019fdgars} and GraphConsis~\cite{liu2020alleviating} construct a single homo-graph based on multiple relations and employ GNNs to aggregate neighborhood information.
GeniePath~\cite{liu2019geniepath} learns convolutional layers and neighbor weights using LSTM and the attention mechanism~\cite{velivckovic2017graph}.
GEM~\cite{liu2018heterogeneous}, SemiGNN~\cite{wang2019semi}, ASA~\cite{wen2020adversary}, and Player2Vec~\cite{zhang2019key} all construct multiple homo-graphs based on node relations in corresponding datasets.
After aggregating neighborhood information with GNNs on each homo-graph, SemiGNN and Player2Vec adopt attention mechanism to aggregate node embeddings across multiple homo-graphs;
while GEM learns weighting parameters for different homo-graphs, and ASA directly sums information from each homo-graph. 
% It is worth noting that SemiGNN adds a graph-based loss to guarantee the local homophily of learned node embeddings.
Player2Vec leverages GCN \& GAT to encode the intra- \& inter-relation neighbor information.
% Some works directly aggregate heterogeneous information in the graph.
% For instance, under an user-review-item heterogeneous graph, 
GAS~\cite{li2019spam} learns unique aggregators for different node types and updates the embeddings of each node types iteratively.

In this paper, CARE-GNN constructs multiple homo-graphs with only one node type like GEM and ASA.
Among the above works, only two works~\cite{wen2020adversary, liu2020alleviating} have noticed the camouflage behaviors of fraudsters.
While \cite{wen2020adversary} only crafts new but inflexible features, and \cite{liu2020alleviating} suffers from unsupervised similarity measures and fixed filtering thresholds.
CARE-GNN remedies those shortcomings by filtering neighbors based on label-aware similarity measures with adaptive filtering thresholds.

% mention attention issue and link to experiment

% our difference and similarity to existing work, our work is graphSAGE, not full batch, no attention

% shrink the introduction of each method

\section{Conclusion}
This paper investigates the camouflage behavior of fraudsters and their negative influence on GNN-based fraud detectors.
To enhance the GNN-based fraud detectors against the feature camouflage and relation camouflage of fraudsters, we propose a label-aware similarity measure and a similarity-aware neighbor selector using reinforcement learning.
Along with two neural modules, we further propose a relation-aware aggregator to maximize the computational utility.
Experiment results on real-world fraud datasets present evidence of fraudster camouflage and demonstrate the effectiveness and efficiency of proposed enhancement modules, especially the reinforcement learning module.

% Future works include investigating the camouflaging behavior and its influence in other fraud problems. The parameter sharing between the similarity measure and the GNN along with the regularization effect of similarity measures is another promising avenue of future research.

\begin{acks}
This work is supported by NSF under grants III-1526499, III-1763325, III-1909323, and CNS-1930941. For any correspondence, please refer to Hao Peng.
\end{acks}

\bibliographystyle{ACM-Reference-Format}
\bibliography{sample-base}

\end{document}